\documentclass[]{aa}

\usepackage{psfig,epsfig,graphicx}
\usepackage{amsmath}
\usepackage{natbib}
\bibpunct{(}{)}{;}{a}{}{,}
\begin{document}
   \title{Contribution of the Disk Emission to the Broad Emission Lines in
AGNs: Two-component model}

   \author{ L. \v C. Popovi\'c\inst{1,2,3}, E. Mediavilla\inst{4},
         E. Bon\inst{1,3}, \and
   D. Ili\'c\inst{5}}

\authorrunning{L. \v C. Popovi\' c et al.}
\titlerunning{Contribution of the Disk Emission to the Broad Emission
Lines in AGNs}

   \offprints{L. \v C. Popovi\'c}

   \institute{
              Astronomical Observatory, Volgina 7, 11160 Belgrade,
Serbia\\
              \email{lpopovic@aob.bg.ac.yu, ebon@aob.bg.ac.yu}
         \and
Astrophysikalisches Institut Potsdam, An der Sternwarte 16, 14482
Potsdam, Germany (Alexander von Humboldt fellow)\\
             \email{lpopovic@aip.de}
\and
Isaac Newton Institute of Chile,
 Yugoslavia Branch
\and
Instituto de  Astrof\'isica de Canarias C/ V\'ia L\`actea,
s/n
E38200 - La Laguna (Tenerife), Spain\\
        \email{emg@ll.iac.es}
\and
Department of Astronomy, Faculty of Mathematics, University of
Belgrade,  Studentski trg 16, 11000  Belgrade, Serbia
\\  \email{dilic@matf.bg.ac.yu}
             }

   \date{Received 03 October 2003; accepted 05 May 2004}

   \abstract{We present an investigation of the
structure of the emission line region in a sample of 12
single-peaked Active Galactic Nuclei (AGNs). Using the high
resolution H$\beta$ and H$\alpha$ line profiles observed with the
Isaac Newton Telescope (La Palma) we study the substructure in the
lines (such as shoulders or bumps) which can indicate a disk or
disk-like emission in  Broad Line Regions (BLRs). Applying
Gaussian analysis we found that both kinds of emission regions,
BLR and NLR, are complex. In this sample the narrow [OIII] lines
are composites of two components; NLR1 which have random
velocities from $\sim$200 to 500 km/s and  systematic velocities
toward the blue from 20 to 350 km/s, and NLR2 with smaller random
velocities ($\sim$100-200 km/s) and a redshift corresponding to
the cosmological one. The BLR also have complex structure and we
apply a two-component model assuming that the line wings
 originate in a very broad line region (VBLR) and the line core in an
intermediate line region (ILR). The VBLR is assumed to be an
accretion disk and the ILR  a spherical emission region.
 The model fits very well the H$\alpha$ and  H$\beta$ line profiles of
the AGNs.
}

   \maketitle
%

\section{Introduction}

The concept of a disk geometry in the Broad Line Region (BLR) is
very attractive because the most widely accepted model for Active
Galactic Nuclei (AGNs) includes  a super massive black hole fed by
an accretion disk. The detection and modeling of some
double-peaked Balmer lines has supported  this idea
\citep{Perez88,
Chen89a,Chen89b,EH94,Rod96,Storchi97,Liv97,Ho00,Shields00,Strateva03,
Berg03,EH03}.
 However, the
fraction of AGNs clearly showing double-peaked profiles is small
 and   statistically insignificant.
On the one hand, the existence of double-peaked lines should not
be required as a necessary  condition for the existence of a disk
geometry in BLRs. Even if the emission in a spectral line comes
from a disk, the parameters of the disk (e.g. inclination) can be
such that one observes single-peaked lines
\citep{Chen89b,Dum90,Koll02,Koll03}. Also, a Keplerian disk with
disk wind can produce single-peaked broad emission lines as
normally seen in most of the AGNs \citep{Mur97}. On the other
hand, taking into account the complexity of emission line regions
of AGNs (see e.g. Sulentic et al. 2000), one might expect that the broad
emission lines are composed of radiation from two or more
kinematically and physically different emission regions, i.e. that
multiple BLR emission components with fundamentally different
velocity distributions are present (see e.g. Romano et al. 1996).
Consequently, one possibility could be that the emission of the
disk is masked by the emission of another emission line region.
Recently, \cite{Pop03} investigated the physical processes in BLRs
using a Boltzmann-plot method, and found that probably 'physical
conditions in regions which contribute to the line core and line
wings are different'. This supports the idea that the broad
optical lines  originate in more than one emission region, i.e.
that the Broad Line Emission Region is complex and composed of at
least two regions. Moreover, \cite{CB96} found that 'the
difference between the Ly$\alpha$ and H$\beta$ full width at zero
intensity (FWZI) values provides additional evidence of an
optically thin very broad line region (VBLR) lying inside an
intermediate line region (ILR) producing the profile cores'.
Consequently we may expect that the VBLR can be formed in a disk
or disk-like emission region. Moreover, recently \cite{Wang03}
investigated the central engines of 37 radio-loud QSOs and found
that their accretion rates suggest that most of the objects
possess standard optically thick, geometrically thin accretion
disks.
 In  fact, recently  \citep{Pop01,Pop02,Pop03a} it was shown
that broad emission lines of at least  three
 AGNs (Akn 120, NGC 3516 and III Zw 2)  can be well fitted with a model
which
has two
 components: (i) an accretion disk and (ii) a region with a geometry
different from a disk.

The aim of this paper is to test the validity of the two-component
model of a BLR which contains an accretion disk and one additional
emission region, i.e. to try to find evidence that suggests that
the disk emission can contribute to the line emission even if the
line profiles are single-peaked. To do this, we observed 12 AGNs
in the H$\alpha$ and H$\beta$ wavelength region with the Isaac
Newton Telescope (Sec. 2, throughout). High resolution spectra
were analyzed first by Gaussian analysis (Sec. 3) and after that
we applied the two-component model for BLRs assuming that it is
composed of a VBLR and an ILR (Sec. 4).

\section{Observations and data reduction}

\begin{table*}
\begin{center}
\caption{The observed AGNs with the coordinates, redshift, central
wavelength and exposition.}
\begin{tabular}{|c|c|c|c|c|c|c|c|}
\hline
Object & RA & DEC & z & Central    & Date of     & Number of & Exposures
\\
name   & h m s    &$\circ$ $'$ ${''}$    &   & wavelength ({\AA}) &
observation & spectra & sec. \\
\hline
\hline
Mrk1040 &  2:28:14.3 & +31:18:40.4 & 0.016652 & 6742.3 & 24-Jan-02 & 4 & 600 \\
 & & & &    4901.6 & 25-Jan-02 & 4 & 1400 \\
\hline
3C120 &  4:33:11.2 & +05:21:27.4 & 0.033010 & 6742.3 & 24-Jan-02 & 4 & 450 \\
 & & & &    5027.9 & 22-Jan-02 & 3 & 900 \\
\hline
NGC 3227 & 10:23:29.6 & +19:52:15.2 & 0.003839 & 6742.3 & 24-Jan-02 & 4 & 480 \\
 & & & &    4899 & 25-Jan-02 & 3 & 600 \\
\hline
PG1116+215 & 11:19:08.7 & +21:19:33.9 & 0.175700 & 5718.5 & 23-Jan-02 & 3 & 1200 \\
 & & & &    7699.1 & 24-Jan-02 & 3 & 500 \\
\hline
NGC4253 & 12:18:26.1 & +29:48:57.5 & 0.012929 & 4475.6 & 23-Jan-02 & 4 & 1200 \\
 & & & &    6742.3 & 24-Jan-02 & 8 & 460, 500 \\
\hline
Mrk110 & 9:25:11.3 & +52:17:29.4 & 0.035291 & 5027.9 & 22-Jan-02 & 3 & 1200, 1500 \\
 & & & &    6742.3 & 24-Jan-02 & 4 & 720 \\
\hline
Mrk 141 & 10:19:10.0 & +63:58:36.6 & 0.041673 & 6744.8 & 24-Jan-02 & 4 & 800 \\
\hline
3C273 & 12:29:08.7 & +02:03:46.7 & 0.158339 & 5669.4 & 25-Jan-02 & 2 & 550 \\
 & & & &    7699.1 & 24-Jan-02 & 3 & 500 \\
\hline
Mrk817 & 14:36:20.5 & +58:48:14.6 & 0.031455 & 4904.3 & 25-Jan-02 & 3 & 550 \\
 & & & &    6742.3 & 24-Jan-02 & 3 & 500 \\
\hline
Mrk493 & 15:59:09.6 & +35:02:21.1 & 0.031328 & 6742.3 & 24-Jan-02 & 3 & 360 \\
\hline
Mrk 841 & 15:04:03.3 & +10:26:48.1 & 0.036422 & 6742.3 & 24-Jan-02 & 3 & 450 \\
 & & & &    4901.6 & 25-Jan-02 & 6 & 300 \\
\hline
PG1211+143 & 12:14:19.4 & +14:03:44.0 & 0.080900 & 5027.9 & 22-Jan-02 & 3 & 600 \\
 & & & &    6742.3 & 24-Jan-02 & 3 & 500 \\
\hline
\end{tabular}
\end{center}
\end{table*}

 It is very important to notice
that the two
 peaks produced by the disk may appear like two bumps in the blue and red
parts of the H$\beta$ and H$\alpha$ line profiles. To find the
substructure connected with disk emission one should obtain the
spectral lines with a relatively high spectral resolution and S/N
ratio.
 We observed with the Isaac Newton Telescope
(INT) 12 AGNs (see Table 1) which
 have been previously observed in the X-ray band (Fe K$\alpha$ line, see
e.g. \cite{Nan97,Sul98})  and
where, according to the X-ray results, 
one can expect that a disk geometry is
present  at least in the X-ray emitting region, i.e. that a disk exists
whose signature  might be observed in
optical lines (emission of the outer part of the disk). The observed AGNs
have no   double-peaked H$\alpha$ and H$\beta$ lines.

The observations were performed  with the 2.5 m INT at La Palma in
the period of 21 - 25 of January 2002.
 The
Intermediate Dispersion Spectrograph (IDS) and the 235 camera
(with chip EEV10) in combination with the R1200Y (for the
H$\alpha$ wavelength region) and R1200B (for the H$\beta$
wavelength region)  gratings were used.
 The list of the observed AGNs with the coordinates, redshift, central
wavelengths and the exposure times are given in Table 1. The
seeing was around $1''.1$ and the slit width was $1''$. The
spectral resolution was $\sim$ 1 \AA . As one can see from Table
1, we observed the H$\alpha$ and H$\beta$ wavelength line region
for all galaxies, except Mrk 141 where only the H$\alpha$ region
was observed. Also, after calibration of the spectra, the H$\beta$
line of Mrk 493 was too weak and the red wing of the 3C 273
H$\alpha$ line was too noisy, so for these two spectra   we use
the low resolution spectra observed with the HST (on Sep 4, 1996
and Jan 31, 1999) with G400 and G750L gratings, respectively.

CuNe and CuAr lamps were used for the wavelength calibration.
Standard reduction  procedures including flat-fielding, wavelength
calibration, spectral response,  and sky subtraction were
performed with the help of the IRAF software package.

 The  software package DIPSO was used for reducing the level of
the local continuum (by using the DIPSO routine 'cdraw 1') fitted
through the dots taken to be on the local continuum (See
Fig. 1).

   \begin{figure}
   \centering
   \includegraphics[angle=0,width=8.8cm]{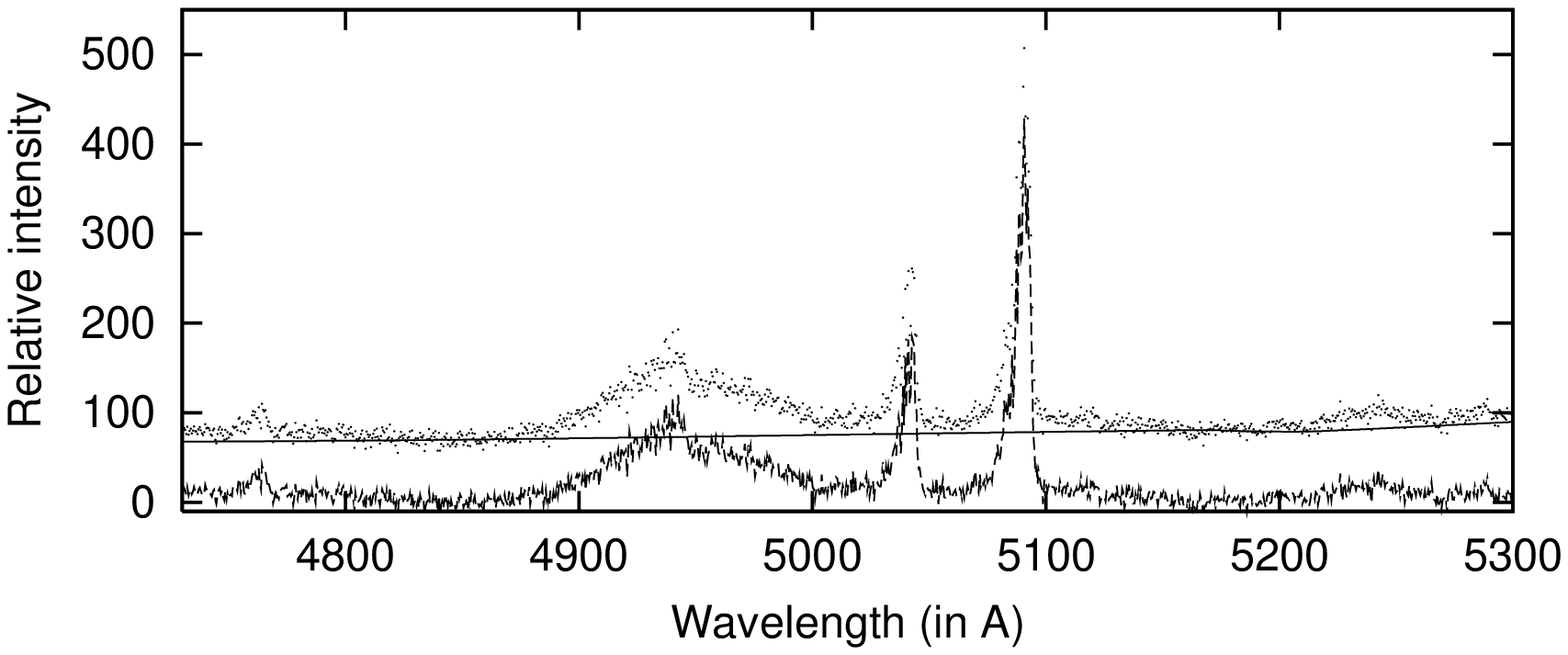}
   \includegraphics[angle=0,width=8.8cm]{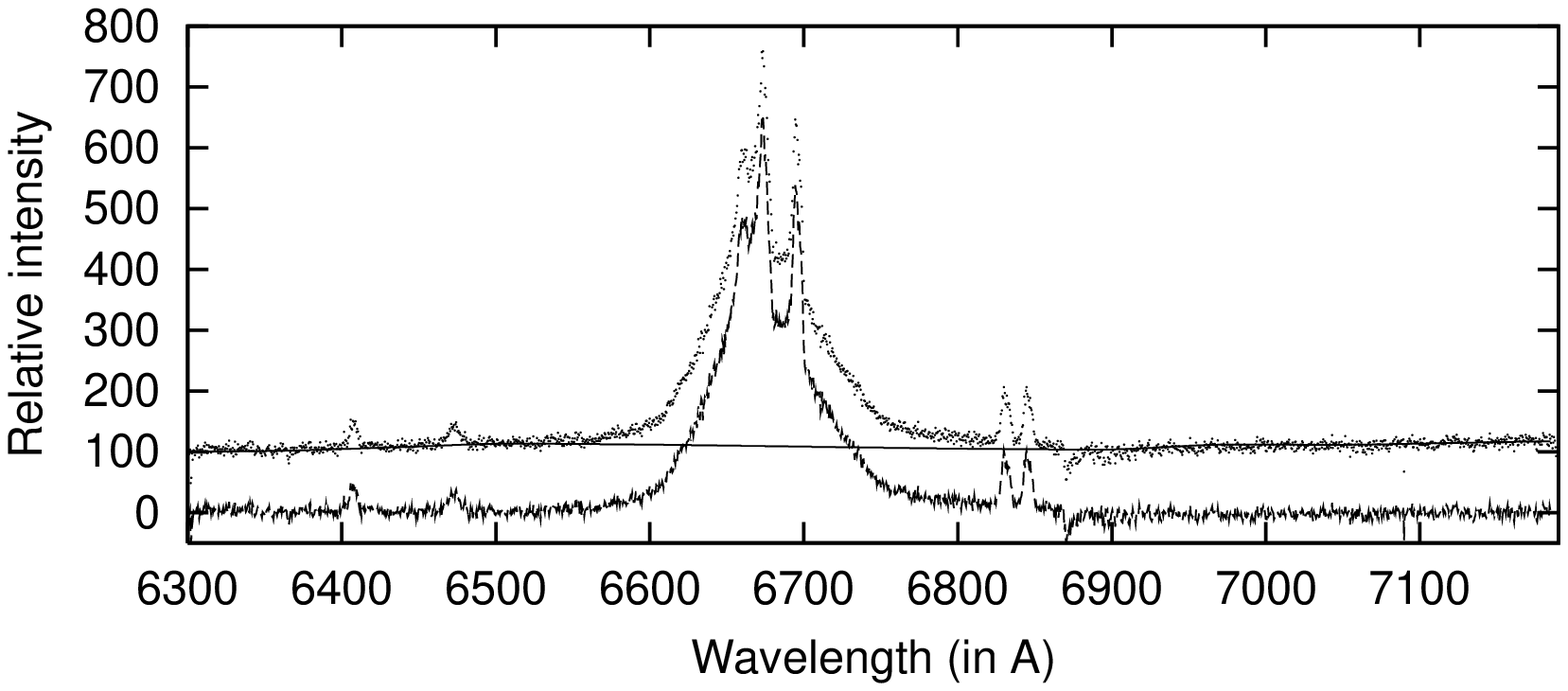}
      \caption{The estimate of the continuum (solid line) in the case of
 Mrk 1040 H$\beta$ (top) and H$\alpha$ (down). Dotted line the observed
spectra and dashed line (below) the spectra after continuum
subtraction.
              }
         \label{FigVibStab}
   \end{figure}

The redshifts of the considered AGNs  were taken from
\cite{Ver00}.

\section{Line profile analysis}

To  analyze  the shape of the H$\beta$ and H$\alpha$ lines, we
first  fitted each line with the sum of Gaussian components. We
used a $\chi^2$ minimalization routine to obtain the best fit
parameters. We also assumed that the narrow emission lines can be
represented by one or more Gaussian components.  In the fitting
procedure,  we looked for the minimal number of Gaussian
components needed to fit the lines. To limit the  number of free
parameters in the fit we set some {\it a priori} constraints
\citep{Pop01,Pop02,Pop03a}:

-- For  the H$\beta$ line
in the fitting procedure we constrained the Gaussian parameters as
follows:

(1) The
three narrow Gaussians representing the
[OIII]$\lambda\lambda 4959,5007$ lines and the narrow H$\beta$ component
are fixed at the same redshift and the Gaussian widths are set
proportional to
their  wavelengths. The Full Width at Half Maximum (FWHM) is connected
with the
width (W)  of the Gaussian profile\footnote{
 given as $\exp{[-(\Delta\lambda/W)^2]}$} by $FWHM=2W\sqrt{\ln 2}$;

(2)   We imposed the intensity ratio of  the two
[OIII]$\lambda\lambda 4959,5007$  lines as 1:3.03 \citep{Wie66}.

(3)  We included  in the fit a red shelf  Fe II template
consisting of  nine Fe II  lines belonging to the multiplets 25,
36 and  42  \citep{Kor92}. We  took  the relative  strength of
these lines from \cite{Kor92}  and assumed  that all  Fe II lines
originate in the  same region,  that is, that all of them have the
same redshift, and widths proportional to  their wavelengths. 
The template contains 9 Fe II lines from 4855.5 \AA\ to 5018.4
\AA, and as one can see from Fig. 2 (e.g. Mrk 493 H$\beta$),
provides a satisfactory fit to the iron lines in the H$\beta$
wavelength region \citep{Kor92,Pop01,Pop02,Pop03a}.  
 In comparison with the list of Fe II lines
 given by \cite{Sipr03},  the Fe II template
used here gives a better fit to the Fe II lines of the present
sample of AGNs.
{ (4) We included  the He I $\lambda\lambda$4921.93,5015.68 \AA\
lines \citep{Veron02}, assuming that they originate in the same
region, i.e. that they have the same redshift and width. Here we
should mention that  the iron lines 4923.9 \AA\ and 5018.4 \AA\
(included in the iron template) are very close to these two He I
lines. Therefore  it is difficult to  find the contribution of
these components. We can obtain a satisfactory fit of the line
profiles without them, but in the case of Mrk 1040, Mrk 110, Mrk
493, Mrk 817 and NGC 4253 the He II lines can contribute to the
red wing of H$\beta$ (see Fig. 2,  full lines in the red wing of
H$\beta$), especially He II$\lambda$ 4923.9 \AA, which tends to be
stronger than He II$\lambda$ 5018.4\AA .}

After a first attempt to fit the lines in the H$\beta$ wavelength region,
we
recognized that the
 [OIII] lines were well fitted only in the central part, but the wings
could not be appropriately fitted with assumption (2), i.e.  the
[OIII]$\lambda\lambda 4959,5007$ lines show extended wings and
cannot be properly fitted by only one Gaussian. We also  notice
that the wings are asymmetrical, with a shallower slope towards
the blue.

To find an acceptable  fit of the [OIII] lines we first included
two more Gaussians which followed constraint (2). However, we
found that a satisfactory fit { in some of the AGNs } can be
obtained only if the intensity ratio of the [OIII]$\lambda$5007
and [OIII]$\lambda$4959 is left as a free parameter. { Since our
main purpose is to find the best fit in order to subtract the
satellite and narrow lines from the H$\beta$ and H$\alpha$ line
profile, we therefore decided to leave as a free parameter the
intensity ratio of the blue-shifted gaussian components of the
[OIII]$\lambda\lambda$4959,5007 lines.} As a rule we find that the
second Gaussian of the [OIII]$\lambda\lambda 4959,5007$ lines is
shifted to the blue and more broadened than the central narrow one
(see Fig. 2).
 { To check our results for the different
components of the [OIII] lines, we subtracted the contributions of
Fe II and He II as well as the H$\beta$ red wing from [OIII]}
using the DIPSO procedure for the continuum (see Fig. 3). Then we
fitted the [OIII] lines  using two Gaussians and the ratio of the
[OIII] lines 1:3.03. In this case we can  fit the [OIII] lines of
some of the AGNs  in an adequate way (e.g. the [OIII] lines of 3C 120, Mrk
817, NGC 3227). There were
 small differences in the residue for Mrk 110, NGC 4253 and PG 1211. But
for the remaining two AGNs (Mrk 1040 and Mrk 841)  we could obtain
a satisfactory fit only with the intensity ratio of
[OIII]$\lambda$5007 and [OIII]$\lambda$4959 left as a free
parameter (see Fig. 4). Indeed, we have measured the fluxes of the
[OIII] lines and found that in these two AGNs the intensity ratio
tends to be smaller than 3.03.  The results of the Gaussian
fitting analysis for the [OIII]$\lambda\lambda 4959,5007$  lines
are presented in Fig. 4, and in Table 2.

-- For the H$\alpha$ line, we  assumed that [NII]$\lambda\lambda
6548,6583$ and the H$\alpha$ narrow component have the same
redshift, and Gaussian  widths proportional to their wavelengths.
Taking into account that the two [NII]$\lambda\lambda 6548,6583$
lines belong to a transition within the same multiplet we assume
an intensity ratio of 1:2.96 (see e.g. Wiese 1966). The fits of
the H$\alpha$ lines are presented in Fig. 5.
 One can expect that other narrow lines in the optical
spectra
have the same shape as the [OIII] ones, but as the
[NII]$\lambda\lambda6548,6583$ lines  are  heavily blended
we are not able to resolve their
 fine structure. Hence, each of them has been fitted by only one
Gaussian. {   On the other hand, we include the He
I$\lambda$6678.15 \AA\ and SiII$\lambda$6371 lines
\citep{Veron02}, and each line has been represented by one
Gaussian. Also, one can expect a contribution of Fe II lines in
the H$\alpha$ wavelength region, but on the bases of the
calculation given by \cite{Sipr03} there are no Fe II
lines\footnote{the nearest lines to H$\alpha$ are Fe
II$\lambda$6456.385 (near the blue wing) and Fe
II$\lambda$7308.065 (near the red wing).} that can significantly
contribute to the H$\alpha$ line profile. Even if any lines are
present  their intensity is probably negligible. It has been shown
(e.g. \cite{Hok87,Sero93}) that AGNs with strong iron emission in
the H$\beta$ wavelength region have a weak emission of Fe II in
the H$\alpha$ band (see e.g. Figs. 1 and 2 in \cite{Hok87}). By
inspection of the  NIST
database\footnote{http://physics.nist.gov/PhysRefData/contents-atomic.html}
in the H$\alpha$  region we found that a possible Fe II
contribution  may come from the permitted lines  $3d^7\ ^2G-3d^6(
^3G) 4s^2G$ (Fe II lines 6700.64 \AA\ and 6873.84 \AA\ ) and
$3d^7\ ^2D2-3d^6( ^3D)4s ^2D$ (lines 6746.53 \AA\ , 6689.41 \AA\
and 6404.615 \AA\ ) as well as from FeII$\lambda$ 6456.387.  In
the fitting procedure we included these lines assuming that they
originate from the same region, i.e. that they have the same width
and redshift. As one can see from Fig. 5 (broad dashed lines),
these lines, as well as the He I and SiII lines, make a negligible
contribution to the H$\alpha$ line profile, i.e. a satisfactory
fit can be obtained without them.}

\subsection{Discussion of the multi-Gaussian analysis}

In Fig. 6 we  present the Gaussian widths  of the  different broad
components versus their centroid velocities (relative to the
systemic one). The different components appear relatively well
separated in this diagram, showing the consistency of the
multi-Gaussian decomposition.  By inspection of  Figs. 2, 4 and 5
as well as of the diagrams in Figs. 6 and 7, we can derive a
number of conclusions concerning the broad line  and the narrow
line components, consequently concerning the BLR and the NLR.

   \begin{figure}
   \centering
   \includegraphics[angle=0,width=8.8cm]{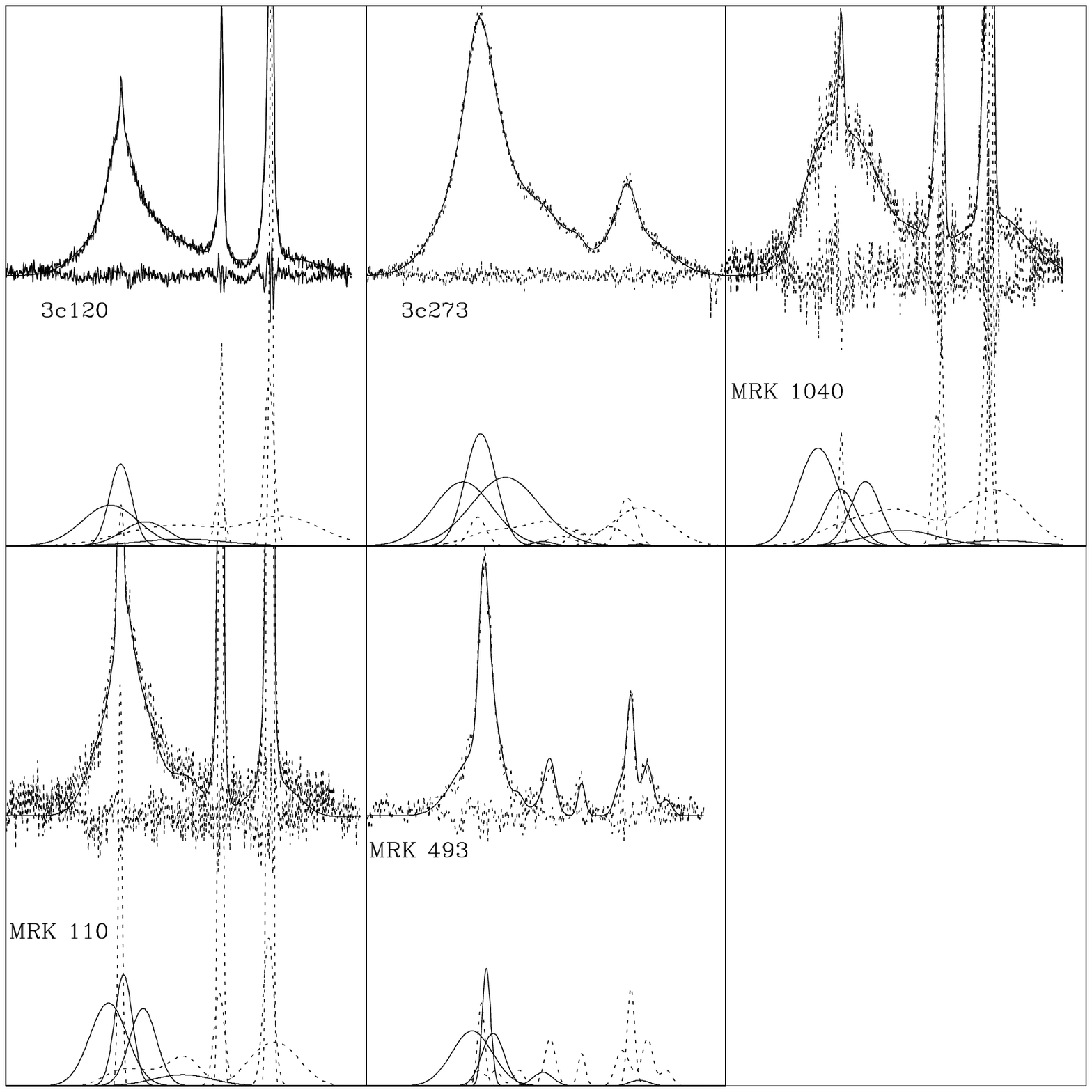}
   \includegraphics[angle=0,width=8.8cm]{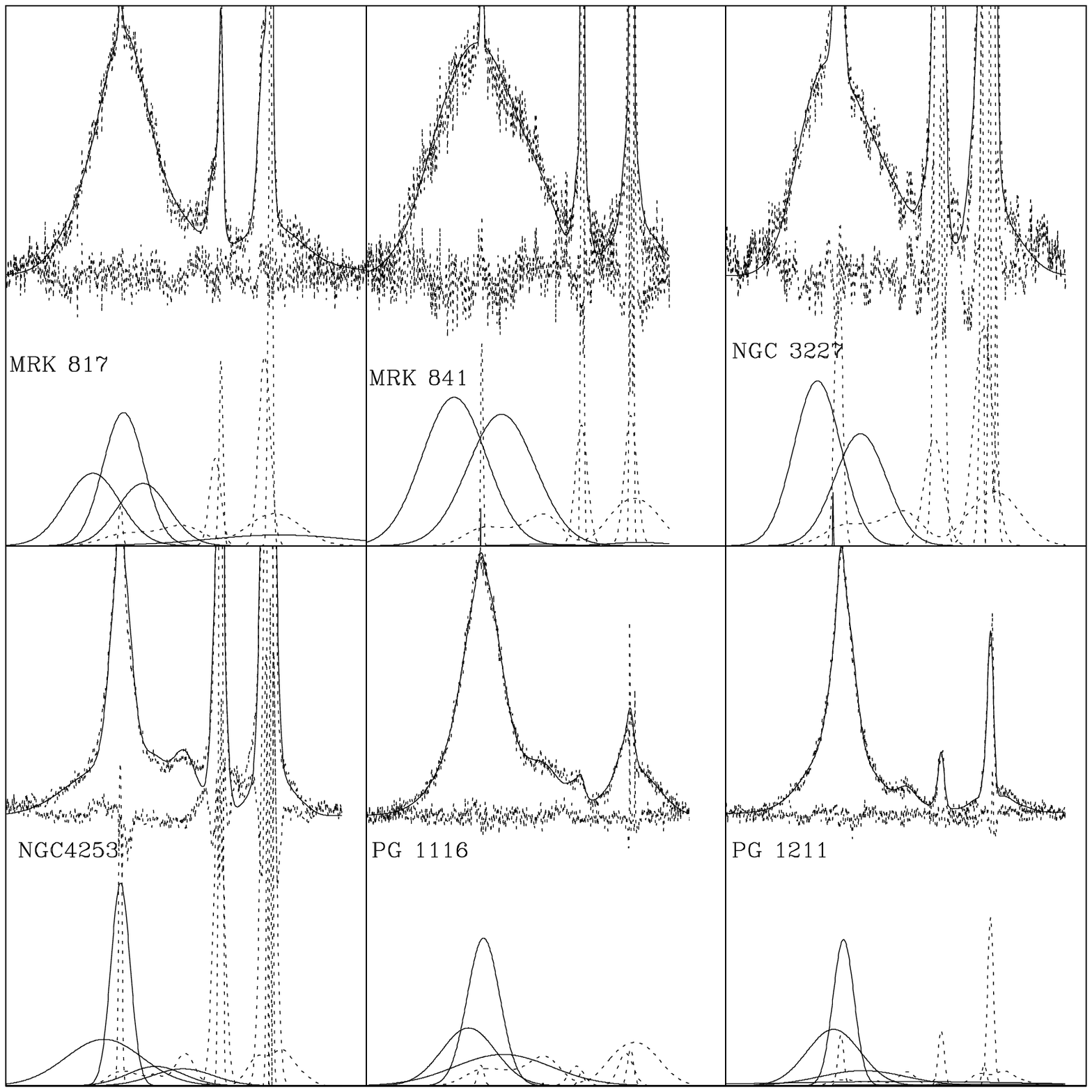}
      \caption{Decomposition of the
H$\beta$ lines  of the observed AGNs.  The dashed lines represent
the observations and the solid lines show the profile
 obtained by Gaussian decomposition.  The Gaussian components are
presented at the bottom.
 The dashed complex lines at the  bottom represent the contribution of
the
Fe II, [OIII] and H$\beta$ narrow  lines
              }
         \label{FigVibStab}
   \end{figure}

\begin{figure}
   \centering
   \includegraphics[angle=0,width=8.8cm]{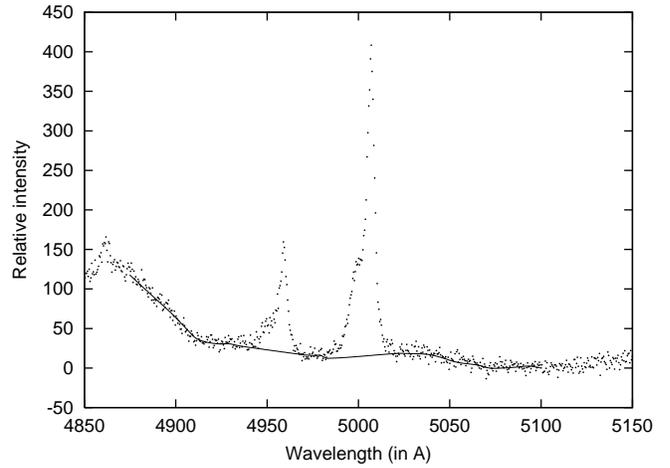}
            \caption{The estimated  contribution of the H$\beta$ wing
and other satellite lines in the [OIII]$\lambda\lambda$4959,5007
line wavelength region (solid line) for Mrk 817.}
         \label{FigVibStab}
   \end{figure}

\begin{figure}
   \centering
   \includegraphics[angle=0,width=8.8cm]{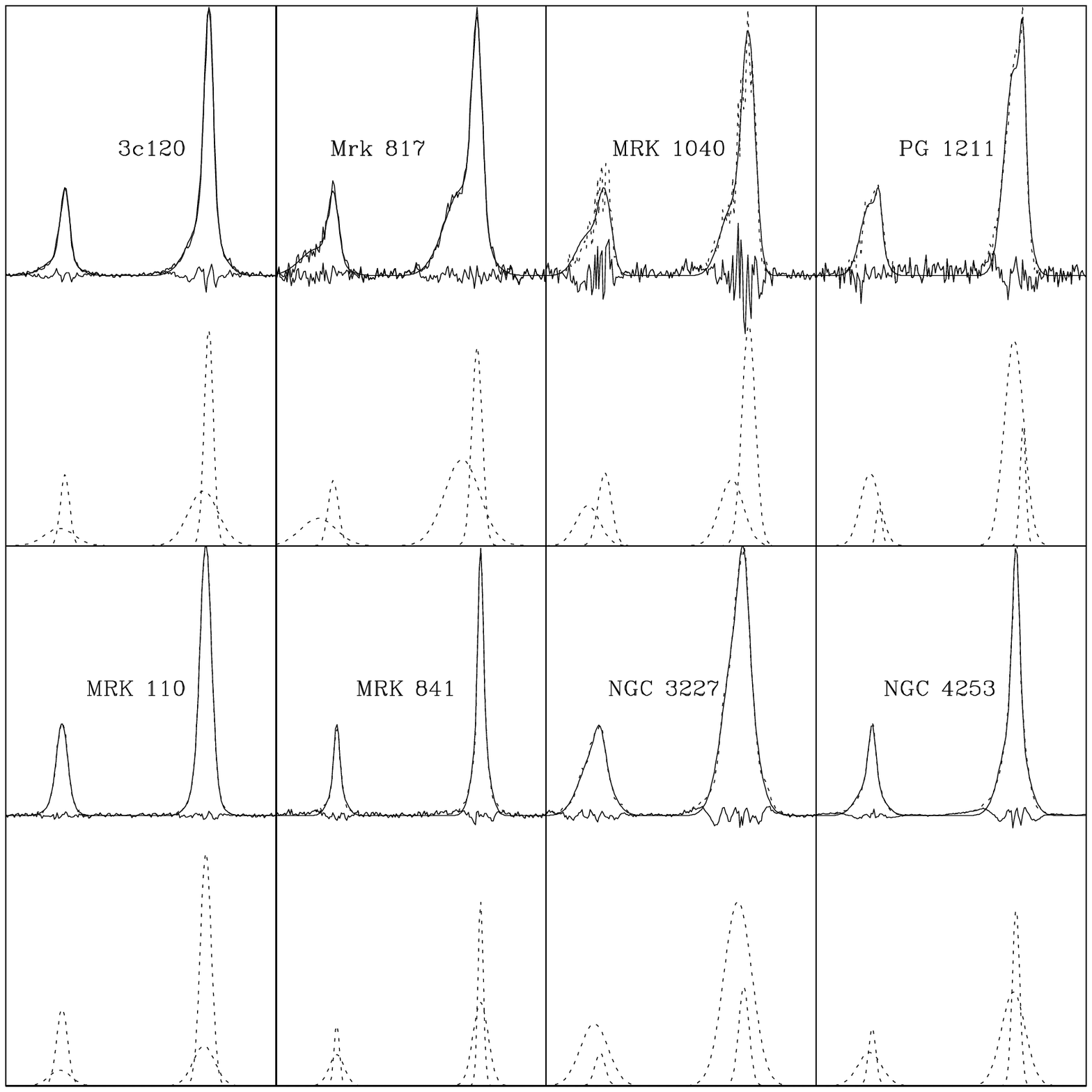}
            \caption{Decomposition of the [OIII] lines
              }
         \label{FigVibStab}
   \end{figure}

   \begin{figure}
   \centering
   \includegraphics[angle=0,width=8.8cm]{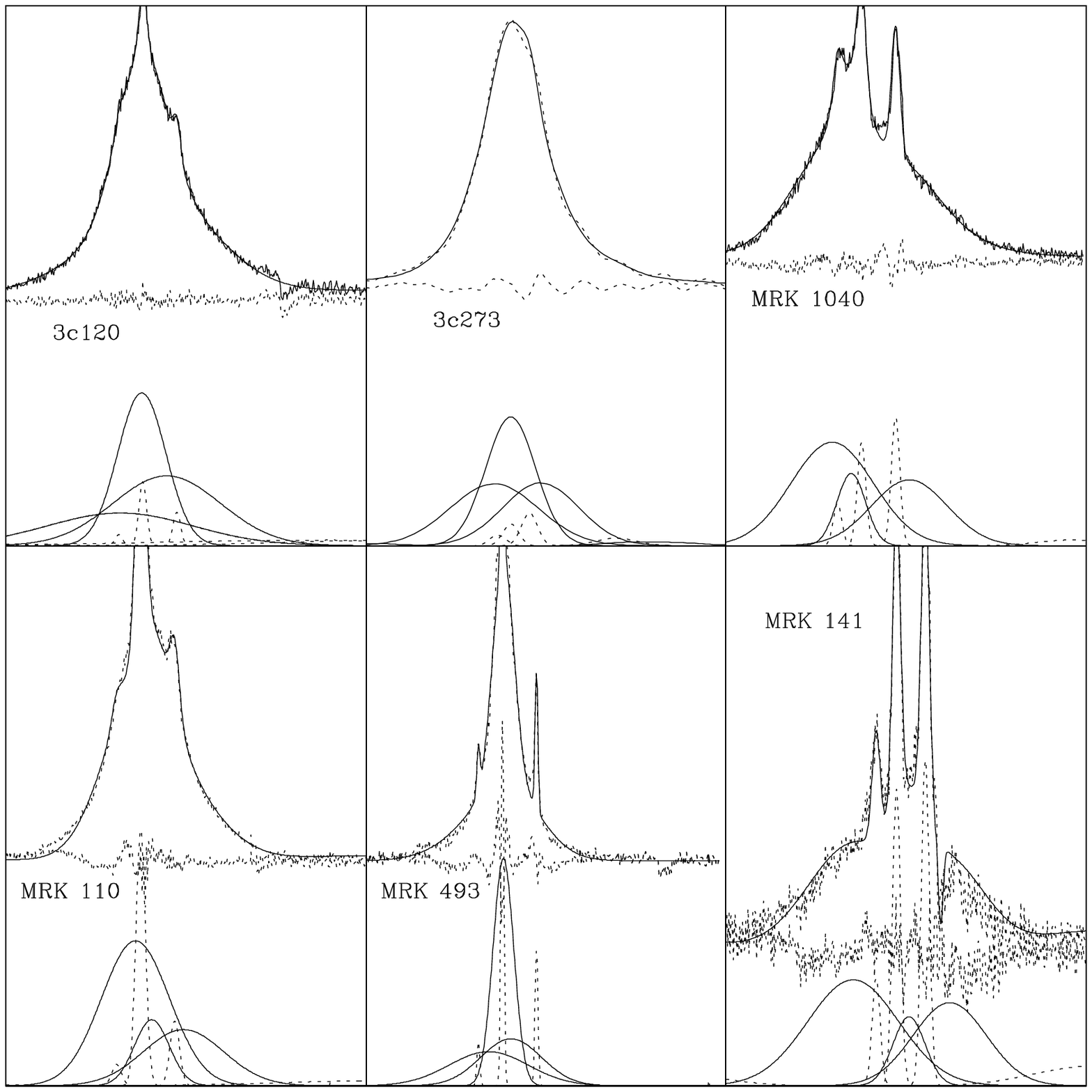}
   \includegraphics[angle=0,width=8.8cm]{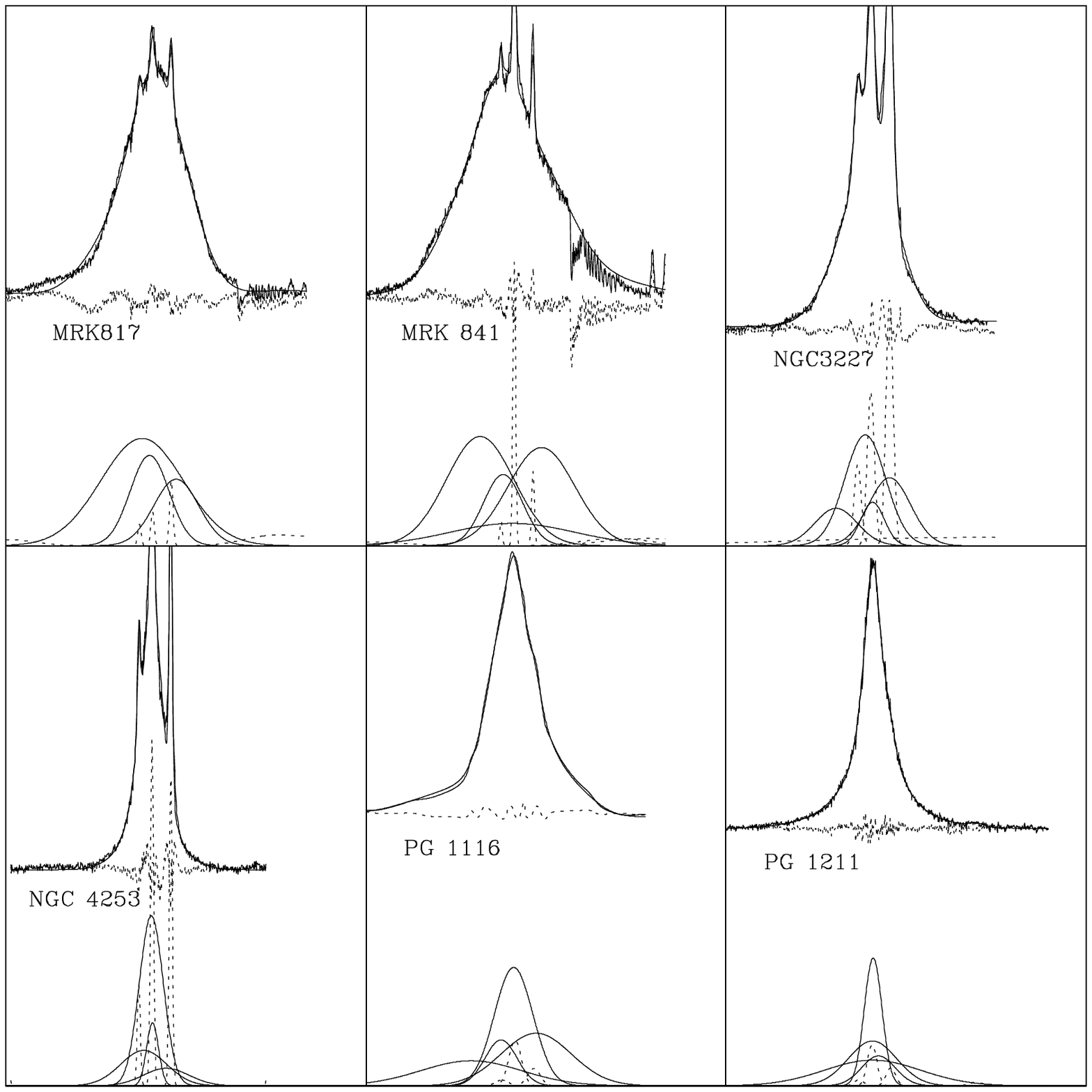}
            \caption{The same as in Fig. 2. but for the H$\alpha$ line. The
narrow dashed lines at the bottom correspond to the narrow H$\alpha$ and
[NII]
lines, the broad dashed lines correspond to the Fe II and He II
contribution.
              }
         \label{FigVibStab}
   \end{figure}

 \begin{figure}
   \centering
   \includegraphics[angle=0,width=8.8cm]{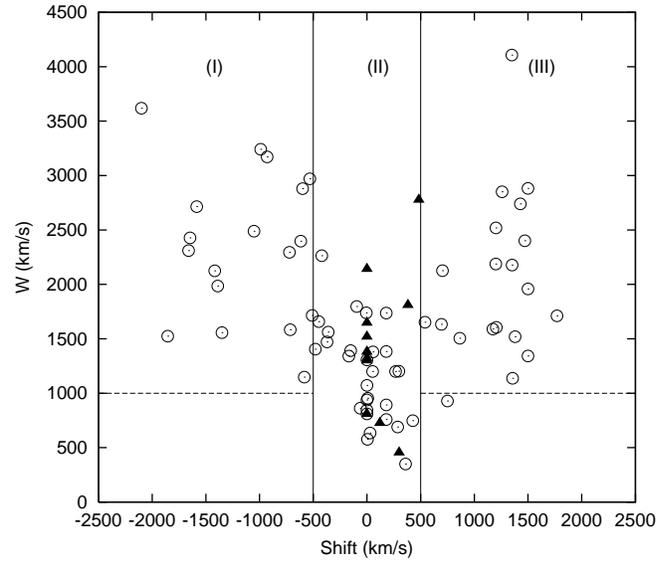}
      \caption{The widths (w) as a function of the inner shifts of the broad
Gaussian components obtained for the H$\alpha$ and H$\beta$ lines
of our sample of AGNs (open circles) and the Fe II template (full
triangles)} \end{figure}

 \begin{figure}
   \centering
   \includegraphics[angle=0,width=8.8cm]{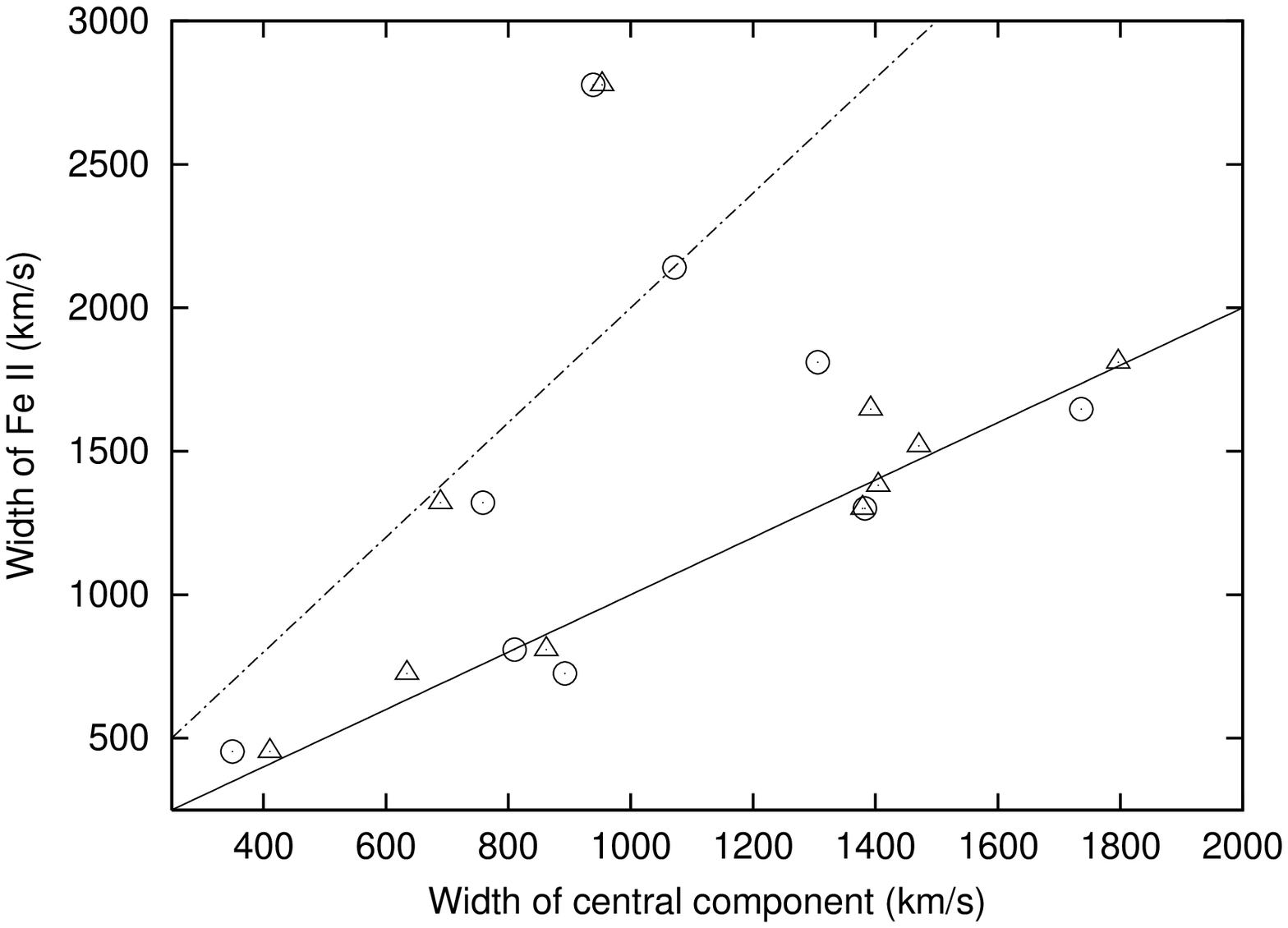}
\caption{ The Gaussian width of the Fe II template ($w_{FeII}$) as
a function of the central broad Gaussian  component width ($w_C$)
of the H$\alpha$ (open triangles) and H$\beta$ (open circles)
line. The solid line corresponds to $w_{FeII}=w_C$ and the dashed
one to $w_{FeII}=2\cdot w_C$. The points at the top, outside of
the region  between the two lines belong to 3C 273.}
 \end{figure}

\subsubsection{The Broad Emission Lines}

Considering only the broad line Gaussian components we can conclude:

(i) that the H$\beta$ and  H$\alpha$ line shapes of the considered
AGNs are very complex, and  usually cannot be described by one
Gaussian, i.e. the Gaussian decomposition indicates a very complex
kinematical structure of the BLRs.

(ii) that the Gaussian decomposition  indicates   the existence of a
central broad component  (part (II) of Fig. 6)
with low random velocities (from 500 to 1500 km/s), and a
redshift consistent with the systemic velocity (between $\pm$ 500 km/s).

(iii) that the  red- and  blue-shifted broad components (part (I)
and (III) in Fig. 6) are often present in the fit. They   tend to
have higher random velocities with higher (positive or negative)
shift. Such a near-symmetrical distribution of the broad red- and
blue-shifted components may indicate that in the considered AGN
sample a fraction of the emission, at least  the wings,
originates in an accretion disk.

{ (iv) The widths of the Fe II lines have different random
velocities (Fig. 6) and a redshift consistent with the systemic
velocity ($\pm500$ km/s). On the other hand, the Fe II random
velocity ($w_{FeII}$) tends to vary linearly with the central
broad component random velocity ($w_{c}$, see Fig. 7),
 with $w_c \le w_{FeII} \le 2w_c$
except in the case of 3C 273 (the two points outside the  region
between these lines in Fig. 7). This may indicate that a
significant fraction of the Fe II emission is created in the
region emitting the core of the H$\alpha$ and H$\beta$ lines.}

\subsubsection{The Narrow Emission Lines}

\begin{table*}
\begin{center}
\caption{The parameters of the Gaussian components  of the [OIII]
lines. Gaussian widths (W) and $\Delta z= z_{NLR2}-z_{NLR1}$ are
given in km/s.  The intensity ratio $R_{NLR1}=I_{5007}/I_{4959}$
of the 'blue component' obtained from the Gaussian fit, the
measured flux ratio $R_F=F_{5007}/F_{4959}$ and central intensity
ratio of the 'blue' and 'central' Gaussian components ($I_B/I_C$)
are also given (columns 2,3 and 6). The data for the objects
denoted with stars should be taken with caution, because their
[OIII] lines are too weak.}
\bigskip
\begin{tabular}{|c|c|c|c|c|c|c|}
\hline

Object  & $W_{NLR1}$ &  $R_{NLR1}$  &$R_F$ & $W_{NLR2}$ &
$\Delta z$&$I_B/I_C$ \\
\hline
\hline

3c120  & 420 &  3.0 &2.9$\pm$0.1 & 130 & -110&0.25 \\
\hline
3c273$*$  & 950  & 2.7 &- & 497 & -200 &0.39 \\
\hline
Mrk 1040& 330 & 1.8& 2.4$\pm$0.1& 190 &-340& 0.30 \\
\hline
Mrk 110 & 300 & 2.6 &3.0$\pm$0.1& 150 & -40&0.17 \\
\hline
Mrk 817 & 510 & 3.0 &3.0$\pm$0.1& 145 & -300&0.43 \\
\hline
Mrk 841 & 215 & 2.6 &2.8$\pm$0.1& 80 & -20& 0.45 \\
\hline
NGC 3227 & 400 & 3.0&3.0$\pm$0.1 & 140 & -120&1.87 \\
\hline
NGC 4253 & 340 & 2.8&3.0$\pm$0.1 & 105 & -55& 0.55 \\
\hline
PG 1116$^*$ & 650 & 1.8& - & 239 & -345& 0.88 \\
\hline
PG 1211 & 270 & 2.8&3.0$\pm$0.1 & 81 & -180& 1.8 \\
\hline

\end{tabular}
\end{center}
\end{table*}

 The NLRs of the AGNs considered  also show a complex structure, and we can
clearly see at least two NLR regions:

(i) the NLR1,
which has an internal random velocity  from $\sim$ 200 to 500 km/s,
and relative approaching velocities  from 20 to 350 km/s with respect to
the
systemic redshift of the observed galaxy; and

(ii) the NLR2 which has an internal random velocity between 100 - 250
km/s, and
a redshift equal to the systemic one of the
corresponding
object.

On the other hand, the intensity ratio of the components from the
NLR2 follows the line  ratio (I(5007):I(4959)=3.03), while for the
NLR1 in 2 AGNs, PG 1116 and Mrk 1040, this ratio is significantly
smaller than 3.03 (see Table 2). Although a slight difference
between the observed and predicted intensity ratio of the [OIII]
lines  might exist \citep{SZ00}, this is certainly due
to the bad S/N for PG1116, and in the case of Mrk 1040, to the
high residues of the fit (see Figs. 2 and 3).

 The difference in shifts and widths of
 [OIII]$\lambda\lambda$4959,5007 between these two NLRs indicate
different kinematical and physical properties.
  Here we note  that the parameters for the NLR1 of 3C 273 and PG
1116+215
given
in Table 2 should be taken with caution because of the low
intensity of the [OIII] lines. We excluded these data in the discussion
above.

The clear tendency of the NLR1 to have a blue-shifted systemic
velocity (though in the case of Mrk 110 and Mrk 841 the blue-shift
is marginal and very close to the cosmological redshift) supports
the idea of a jet geometry of the NLR (see e.g. Dopita et
al. 2003). {
In this case the receding jet component in the [OIII] lines might
be  obscured or absorbed by the host galaxy, so one sees only the
outflowing gas from  the closer part of the jet.}
 { Many papers have been
devoted to  the radial velocity difference between narrow and
broad lines (starting e.g. from Gaskell 1982), especially the
velocity difference between the [OIII] lines and  H$\beta$ that
indicates a jet geometry of the NLR or of part of the NLR (see
e.g. Bennert 2002, Zamanov 2002). On the other hand,  2D
spectroscopy
clearly shows that  AGN NLRs have a complex structure (see e.g.
Arribas 1997).
 The tendency for the [OIII] lines to be blue-shifted
relative to the H$\beta$ suggests that they are associated
with a
high-ionization outflow originating in these highly accreting sources
(see e.g. Zamanov 2002).}

\section{Two-component model for BLR}

Here we  apply the two-component model in  modeling the broad line
shapes of 12 AGNs, where one component is the disk or disk-like
region.
 We start from the paper of
\cite{CB96}, who investigated the combined ultraviolet and optical
spectra of 48 QSOs and Seyfert 1 galaxies in the redshift range
0.034-0.774. They found a statistically significant difference
between the FWZI distributions of the Ly$\alpha$ and H$\beta$
lines. The difference between the Ly$\alpha$ and H$\beta$ FWZI
values provides additional evidence for an optically thin VBLR
(which might be a disk or disk-like region) which contributes to
the line wings. It is located inside an ILR which produces the
profile cores. Also, they found { relative weakness of the
correlations between the UV profile asymmetries and widths and
those of H$\beta$ line}. This suggests a stratified  structure of
the BLR, consistent with the variability studies of Seyfert 1
galaxies (see e.g. Kollatschny 2003). The smaller average FWHM
values
of the UV lines compared to H$\beta$ indicate that the ILR emission make a
higher contribution to the UV lines,
whereas in the Balmer lines the VBLR component is more dominant. This is
also the case
in well known AGNs with double-peaked Balmer lines, which usually
show a single-peaked  Ly$\alpha$ line (see e.g. the case of Arp
102B, Halpern et al. 1996).

The wings of the broad H$\alpha$ emission line in the spectra of a
large sample of AGNs (around 100 spectra) were investigated by
\cite{Rom96}. They found an indication of multiple BLR emission.
Moreover, recently \cite{Pop03} investigated the physical
conditions in BLRs finding an  indication that  BLRs are
 complex and that
physical conditions of the regions which contribute to the line core and
the line wings are different.

Though a two-component model can probably also be represented by
other geometries, we choose the one with:  a disk giving the wings
of the lines, and a spherical component giving the core of the
lines.

\subsection{Theoretical point}

For the disk we  use the Keplerian relativistic model of
Chen
\& Halpern (1989). The emissivity of the disk as a function of radius,
$R$, is given by
$\epsilon=\epsilon_0R^{-p}.$
{ Generally,  when trying to
  fit the double-peaked line profiles by disk emission one should  leave
this index as a free parameter. However, we have to take  into
account two facts: (1) we have here single-peaked lines, i.e. the
profile coming from the disk is not {\it a priori} well defined;
(2) we will be using a two-component model which includes more
parameters than a disk-only model. We should therefore include
some constraints.} Since  the illumination is due to a point
source radiating isotropically, located at the center of the disk,
the flux in the outer disk at different radii should vary as
$r^{-3}$ \citep{EH94}. { We note here that  this is indeed the way
the incident  flux varies, but not necessarily the way in which
the lines respond to it \citep{Dum90,RB99,EH03}.
However,
 the power index $p\approx 3$ can be adopted  as a reasonable
prescription at least for H$\alpha$  \citep{EH03}. In our case, as
one can see in Fig. 8, the shapes of H$\alpha$ and H$\beta$  in
our sample are  practically the same, so we can fit an averaged
profile (see Sect. 4.2). { Therefore  we will start our fitting
procedure by imposing the constraint $p = 3$,
 but later we will change this parameter and use
other constraints (see Sect. 4.2)}.

 We express the disk dimension
in gravitational radii ($R_g=GM/c^2$, $G$ being the gravitational
constant,
$M$  the mass of the central black hole, and $c$  the velocity of light).
The  local broadening parameter ($\sigma$) and shift ($z_l$) within the
disk
have been taken into account as in \cite{Chen89b}, i.e.
the $\delta$ function has been replaced by a Gaussian function:
$$\delta\to \exp{{{(\lambda-\lambda_0-z_l)^2}\over{2\sigma^2}}}$$
where $z_l$ and $\sigma$ are the local shift and the broadening
parameter of the disk emission, respectively.

   \begin{figure}
   \centering
   \includegraphics[angle=0,width=8.8cm]{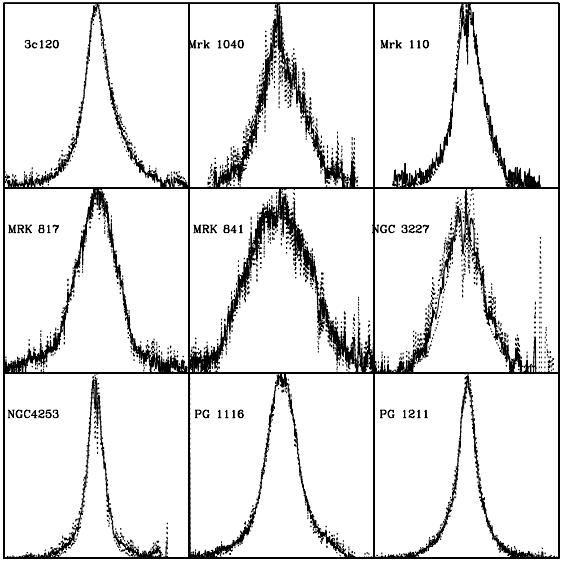}
      \caption{The comparison of H$\beta$  and H$\alpha$
(dashed lines) profiles with an averaged one (solid line).
              }
         \label{FigVibStab}
   \end{figure}

   \begin{figure}
   \centering
   \includegraphics[angle=0,width=4.2cm]{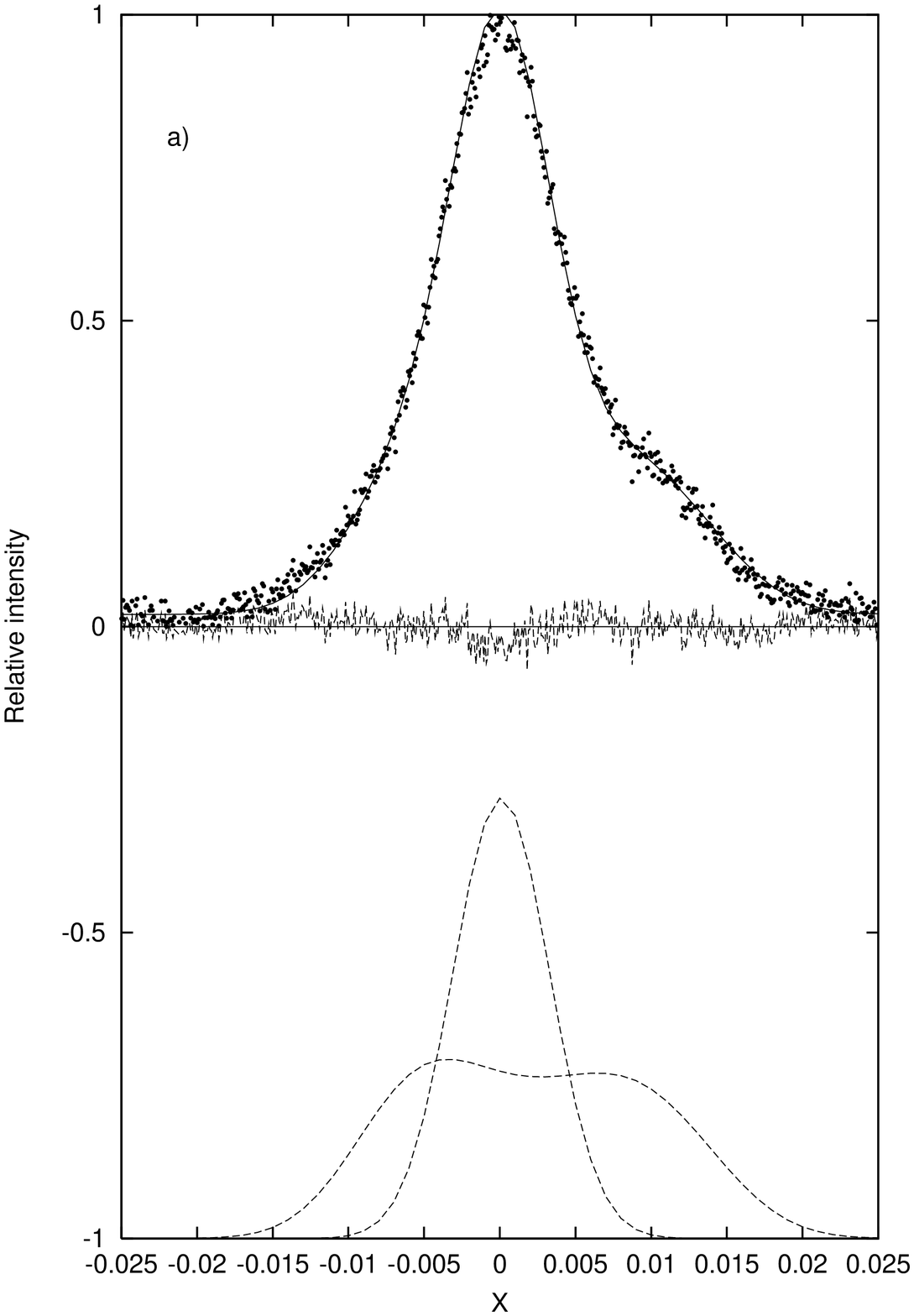}
  \includegraphics[angle=0,width=4.2cm]{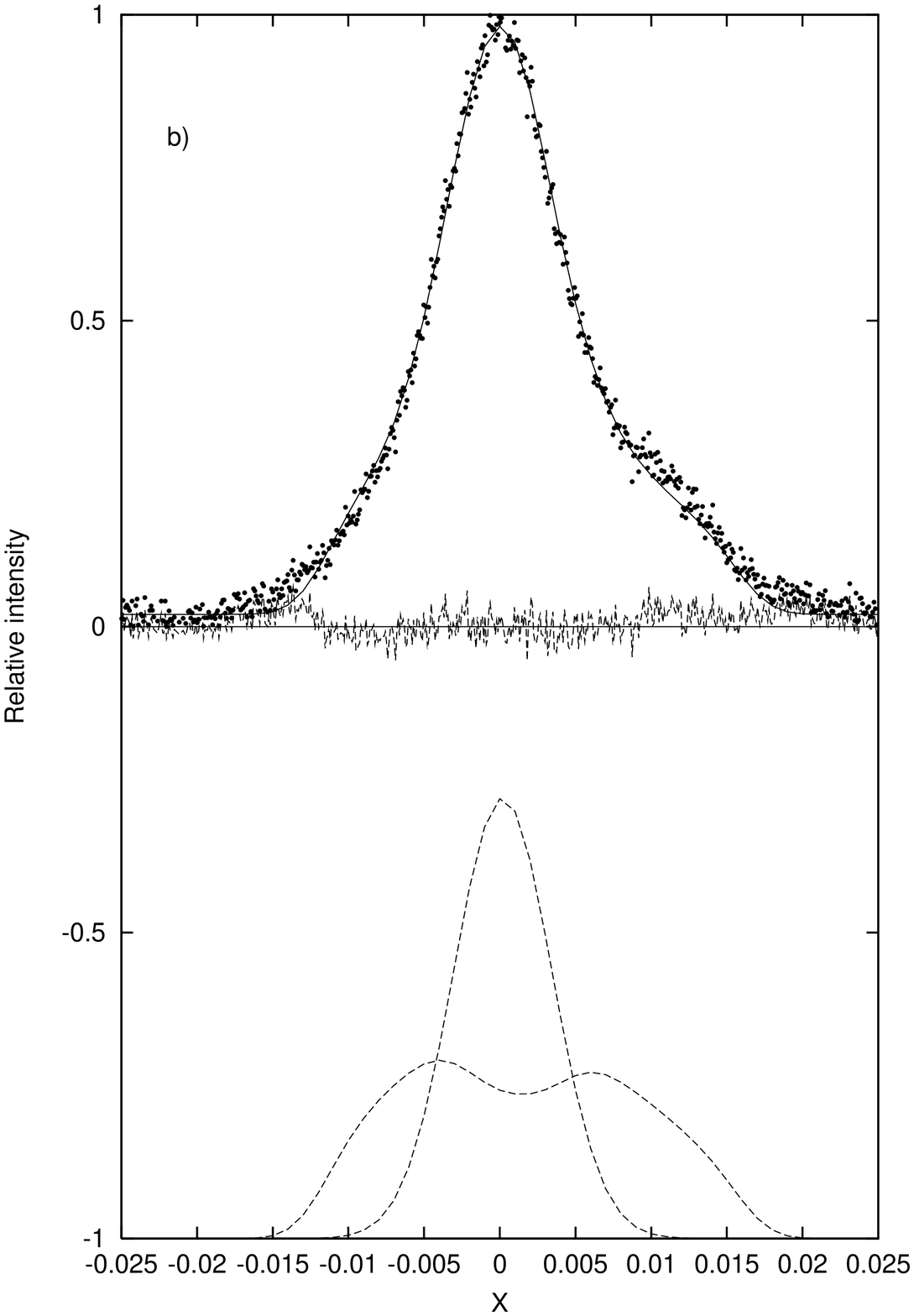}
      \caption{Two fits of 3C 273 with the two-component model.
The  disk parameters are: a) $i=14^\circ$, $R_{inn}$=400 $R_g$,
$R_{out}=$1420 $R_g$, W$_d$=1620 km/s, p=3.0 (W$_G$=1350 km/s);
b) $i=29^\circ$,
$R_{inn}$=1250 $R_g$, $R_{out}=$15000  $R_g$, W$_d$=700 km/s,  p=2.8
(W$_G$=1380 km/s)              }
         \label{FigVibStab}
   \end{figure}

On the other hand, we assume that the kinematics of the additional
emission
region can be described as the emission of a spherical  region
with
{ an isotropic velocity distribution}, i.e. with
a local broadening $w_G$ and shift $z_G$. Consequently, the emission
line
profile
can be described by a Gaussian function.
The whole line profile can be described by the relation:

$$I(\lambda)=I_{AD}(\lambda)+I_G(\lambda)$$
where $I_{AD}(\lambda)$,  $I_G(\lambda)$ are the  emissions of the
relativistic accretion disk and of  an additional  region,
respectively.

\subsection{Fitting procedure}

Before performing the fitting we 'cleaned' the spectra by
subtracting (i) from H$\beta$ the narrow H$\beta$ line, the narrow
[OIII] lines, the He I lines, and the Fe II template; (ii) from
H$\alpha$ the narrow H$\alpha$ and [NII] lines. Furthermore, we
normalized the intensities of H$\alpha$ and H$\beta$ to unity and
converted the wavelength into a velocity scale:  $\lambda \to
X=(\lambda-\lambda_0)/\lambda_0$. These conversions allowed us to
compare the H$\alpha$ and H$\beta$ high resolution profile. We
found that in the AGNs of the sample the H$\alpha$ and H$\beta$
have similar profiles (see Fig. 8). That concept supports the case
that both  lines are formed in the same emission region. First we
found an averaged line profile from the H$\alpha$ and H$\beta$
lines from each AGN (here we should note that for 3C 273 only the
high resolution H$\beta$, and for Mrk 493 and Mrk 141 only
H$\alpha$ were used). The averaged profile for each AGN was fitted
with the above described two-component model.

When a chi-square minimization including all the parameters  was
attempted, it was found that the results are very dependent on the
initial values given to the parameters. { As mentioned above, the
reason for this is that we apply a two-component model to
single-peaked lines, so
 the number of free parameters is
too large.}
 To overcome this problem we have { to use the additional constraint that
the disc component fits the line wings, and the spherical
component the line core.} With this aim,
 we tried several values for the
inclination.  The starting point for the inclination can be the
relative displacement of red- and blue-shifted Gaussians, $\Delta
z$. We use the  empirical relation given by \cite{Smak81} (see
also Popovi\'c 2002,2003}). If we assume that a disk (or a
disk-like) region exists, we can roughly estimate the parameters
of the disk using the results of Gaussian analysis and the
relationship (see Popovi\'c 2002 for more details) $\sin i\approx
{\Delta z\sqrt{2\cdot R_{\rm out}}},$ where $i$ is the inclination
of the disk, and $R_{\rm out}$ is the outer radius given in
gravitational radii.  Taking into account that $\sin i\leq1$, we
can estimate the maximal outer radius. From Fig. 6 one can
estimate that
 $\sim$ 0.00033 ($\sim$1000 km/s) $\le\Delta z\le$ 0.013 ($\sim$4000
km/s), thus the maximal outer radius may be in the interval $\sim$
a few $10^3R_g$ to $\sim$ a few $10^5R_g$ for the highest
inclination. On the other hand,   from previous investigations
\citep{Chen89a,EH94,Era03} it was found  that the outer radius of
the disk has typical dimensions of
 $\sim$ a few $10^3 R_{g}$; in that case, the inclination for the
sample is in the interval  $i\sim 5^\circ-25^\circ$.

{ As we noted in Sect. 4.1, first we fixed $p=3$, and chose
different values of the inclination}.
 Then we found the best fit by eye and after that we used a { 'fixed'
inclination such that other parameters can change more (with
greater step) than the inclination\footnote{we use the SIMPLEX
method for fitting that allows us to define  the different
magnitudes of 'reflections' (the steps to change the
parameters).}. After that we } performed a chi-square fitting of
the  parameters starting from suitable initial values. { Note that
this is not an 'orthodox' chi-square fitting procedure, but in
this case the main point is not only to find the best fit, as we
did with the Gaussian fitting procedure, but also to  try to
explain the line profile with our two-component model.
Consequently, we found that}
 the fit of the wings of the BELs  strongly restricts the value of the inner
 radius and the  "local"
 broadening, i.e. the random velocity of emission gas in the disk. {
 We found that the inclinations of the disk are relatively small,
between $5^\circ$ and $15^\circ$. One should have expected higher
inclinations in the sample (e.g. on average 30$^\circ$).

Therefore we decided to choose the inclination in the range $i\ge
25^\circ$, leaving $p$ as a free parameter, and looking for the
maximal inclinations. We obtained
 reasonably good fits for the lines of more than half of the sample,
for $p< 3$ and higher values of the inner and outer accretion disk
radii (see Fig. 9ab). Note that  low values of the emissivity
index, $p$, were obtained also by
 \cite{EH94} and \cite{Era03} when fitting double peaked
lines. Consequently, we did fitting tests for different fixed
values of $p$ in the range from 1.0 to 3.0. Also, in this case we
can find acceptable fit for some  AGNs.
 As an example, in Fig. 9 one can see  two fits of the 3C 273
line with different parameters. The best fit corresponds to Fig. 9a, but
 the fit in Fig. 9b is
also reasonable.
 From these fitting tests (without any constraints
for the disk parameters) we are able only to give rough estimates
of the disk parameters  (see Table 3). We should mention here that
for lines with the smallest asymmetry it was more difficult to
estimate the parameters. Consequently,  the estimated range of the
parameters is higher (see Table 3, e.g. Mrk 110).}

\begin{table*} \caption{The parameters of the disk: z$_{\rm l}$ is the
shift and $W_l=\sqrt{2}\sigma$ is the  Gaussian broadening term
from the disk which is a measure of the random velocity in the
disk, R$_{\rm inn}$ is the inner radius, R$_{\rm out}$ is the
outer radius. z$_{\rm G}$ and W$_{\rm G}$ represent the parameters
of the Gaussian component.}

\vbox{\tabskip=0pt   \offinterlineskip  \def\podvuci{\noalign{\hrule}}
\def\razmak{\noalign{\vskip.1cm}}     \halign    to    \hsize{\strut#&
\vrule#\tabskip=0pt  plus   10pt  minus5pt  &
 \hfil#\hfil&  \vrule#&
 \hfil#\hfil&  \vrule#&
\hfil#\hfil&  \vrule#&
\hfil#\hfil&  \vrule#&
 \hfil#\hfil&  \vrule#&
\hfil#\hfil&  \vrule#&
\hfil#\hfil&  \vrule#&
 \hfil#\hfil&  \vrule#&  \hfil#& \vrule #
  \tabskip=0pt\cr\podvuci
&& Object &&$i$&& z$^{\rm min,max}_{\rm l}$  &&$W^{\rm min,max}_l$
(km/s) && R$^{min}_{\rm inn}\
(R_g)$
&&R$^{\rm max}_{\rm out}\ (R_g)$  && z$^{\rm min,max}_{\rm G}$ && W$_{\rm
G}$
(km/s)&&
$p^{\rm min}$&\cr
\podvuci\razmak\podvuci
&& 3C 120 && 8-30 && -300,+300 &&  1050,1500 && 350 &&  20000 && +30,+300
&& 900$\pm$150 &&2.0 &\cr
&& 3C 273 && 12-30$<$ && -30,+300 && 690,1760 && 400  && 15400 && +30,+60
&& 1380$\pm$150&& 2.3 &\cr
&&MRK 1040&& 5-27$<$ && -250,+300  && 800,1400 && 100&& 18000 && 0$\pm$30
&&
500$\pm$200&& 1.3 &\cr
&&MRK 110&& 7-50  &&-320,+300  && 450,1250  && 400&& 49000 && +150$\pm$30
&&
960$\pm$50&&1.7 &\cr
&&MRK 141 &&12-33 && -630,-450 && 700,1500 && 300 && 10000  && +200,+300
&& 1620$\pm$100&&2.1 &\cr
&&MRK 493 &&5-30$<$ &&  -480,+60 && 360,560 && 600 && 124000  &&
+60$\pm$30 &&
360$\pm$50&&1.8 &\cr
&&MRK 817 && 12-35 && -450,+300&& 850,1200 && 140 && 14000  && 0,+130  &&
1550$\pm$100&&1.8 &\cr
&&MRK 841 &&15-50 &&-750,-150 && 1070,1800 && 450 && 27400  && -300$\pm$30
&&
1500$\pm$100&&2.1 &\cr
&&NGC 3227&&12-34 && -780,-300 && 900,1550 && 350 && 12000  && -300,300 &&
1500$\pm$100&&2.1 &\cr
&&NGC 4253&&5-25$<$ && -630,-90 && 280,850 && 500 && 69500  && -90,-30  &&
550$\pm$50&&2.0 &\cr
&&PG 1116 &&8-30$<$ &&  -450,0 && 1100,1800 && 500 && 15800  && 0,+90   &&
1400$\pm$250 &&2.2 &\cr
&&PG 1211 &&8-30 && -660,0 && 540,1100 && 600 && 67400  && +90$\pm$30  &&
600$\pm$300&& 1.9 &\cr
\podvuci}}
\end{table*}

\section{Results and Discussion}

{ As one can see from Fig. 9, the line profiles can be well fitted
with the two-component model, but some of the parameters (e.g. the
emissivity index, the inclination, the inner and outer radii) are
not constrained. It is therefore not possible  to find a unique
solution for  the model. For this,
 one should arbitrarily constrain at least
one of the disk parameters\footnote{ e.g. the dimensions of the
BLR obtained  from reverberation studies might be used as a
constraint.}.   However, it is obvious that in the  AGNs of the
sample, the shape of the line wings indicates  radial motion,
which may be caused by a disk-like geometry. In any case, the
fitting tests described above  allow us to obtain rough estimates
of the kinematical parameters of the two-component model (Table
3). In Table 3 we present the estimated range of inclinations (i)
and the minimal emissivity index ($p^{\rm  min}$), minimal and
maximal value for shifts and widths of the Gaussian broadening
term from the disk (z$^{\rm min,max}_{l}$, W$^{\rm min,max}_l$),
the shifts and widths of the Gaussian component (Z$_G$,W$_G$), and
estimates for the minimal inner radius (R$^{\rm  min}_{\rm inn}$)
and maximal outer radius (R$^{\rm  max}_{\rm out}$). }

   \begin{figure}
   \centering
   \includegraphics[angle=0,width=8.8cm]{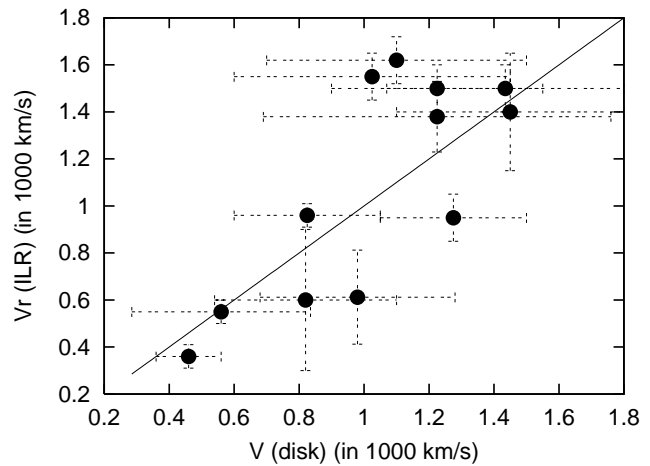}
            \caption{The  random velocities of a spherical region (ILR) as
a function of the local random disk velocities. The dashed line
represents the function $V_{ILR}=V_{Disk}$ km/s}.
         \label{FigVibStab}
   \end{figure}

{
 Concerning the  disk  we can  point out the following:
  (i) The maximal outer radius is in the range $10^4$ to $10^5\
R_g$ (ii)  The local random  velocities in the disk are different
from object to object and they are in the broad range from $\sim$
300 km/s to 1700 km/s; so are the local redshifts  ($z_{\rm l}\sim
-780$ to $+450$ km/s). (iii) The minimal inner radius of the
emitting disk is in the range from $\sim$ 100 to 600 R$_g$, (iv)
The inclinations are  $5^\circ\leq i<50^\circ$, and in about
 half
of the AGNs  $i<30^\circ$. Such values of the inclinations support
the idea  that we
 more frequently
observe the Sy 1 type of AGNs at   face-on
inclinations \citep{RB99}.}

Concerning the spherical emission region we can point out that: i)
the redshifts are within 300 km/sec of the cosmological value;
 (ii) the random velocities in this
region are also different for different objects, they are in the
range from $\sim$400 to 1600 km/s.

Particularly, it is interesting to see the correlation between
local broadening in the two regions presented  in Fig. 10. The
local random velocities in the disk (averaged value from Table 3)
are well correlated with those of the spherical region. This can
be explained if this region   originates from an accretion disk
wind, which is created through several disturbances in the disk
capable of producing shocks (e.g. Bondi-Hoyl flow, stellar
wind-wind collision, and turbulences).
 Recently, \cite{From01}  described a scenario for the
formation of a part of the BLR caused by shocks in the accretion
disk; this may 
also create an ILR. 

We should mention that besides a disk (or a disk-like region) or
spiral shock waves within a disk \citep{Chen89a,Chen89b}, other
geometries may cause the same kinds of substructure in the line
profiles: i) emission from the oppositely-directed sides of a
bipolar outflow \citep{Zhen90,Zhen91}; ii) emission from a
spherical system of clouds in randomly inclined Keplerian orbits
illuminated anisotropically from the center \citep{GW96}; and iii)
emission from a binary black hole system \citep{Gas83,Gas96}. In
any case  one should consider a two-component model with an ILR
contributing to the broad line cores  and one additional emitting
region contributing to the broad line wings. Recent investigations
(see e.g. Wang et al. 2003, Eracleous \& Halpern 2003)
 have shown that the disk geometry for VBLR may be accepted as a
reality. Moreover, \cite{EH03} found  that  the disk emission is
more successful  not only in explaining double-peaked line
profiles but also in interpreting   the other spectroscopic
properties of AGNs presenting these  double-peaked Balmer lines.

\section{Conclusions}

We observed  12 AGNs  with INT in order to obtain  high-resolution
spectra of the H$\beta$ and H$\alpha$ lines that can be used for
modeling BLRs. First we applied Gaussian analysis to the complex
H$\beta$ and H$\alpha$ lines, from which we can conclude: (i) both
the BLR and the NLR are complex, (ii) the [OIII] narrow lines can
be fitted satisfactory only with two Gaussians, one shifted toward
blue, which may indicate the existence of an outflow, (iii) the {
different parameters of
 these two Gaussians indicate that they originate in two NLR regions
having different physical properties}, (iv) the broad lines also
show a complex structure, and they can be decomposed into three
broad Gaussians - one red-, one blue-shifted and one at the
systemic redshift.

We adopt a two-component model, which comprises a VBLR and an ILR.
We identify  the VBLR with  an accretion disk which contributes to
the line wings. The cores of the lines originate in the ILR which
is assumed to have a spherical geometry. This two-component model
has been applied to the observed line profiles and we can conclude
that : { (i) The model can very well fit the observed line
profiles, but it is very hard to obtain the disk parameters
without imposing at least one constraint  because of the large
number of parameters and the lack of two peaks in the line
profiles. They can be only roughly estimated using  fitting tests
(see Table 3).
 (ii) The
random velocities in the spherical emission region and the  random
velocities in the disk  are similar. This indicates that these two
regions are linked through a common process, such as a wind
produced by the disk.

To find constraints for the model parameters further
investigations are needed.

\begin{acknowledgements}
The work was supported by the Ministry of Science, Technologies
and Development of Serbia through the project "Astrophysical
Spectroscopy of Extragalactic Objects" (L. \v C. P., E.B, \&
D.I.), the IAC through the project P6/88 "Relativistic and
Theoretical Astrophysics" (E.M. \& L. \v C. P.) and the Alexander
von Humboldt Foundation through the program for foreign scholars
(L. \v C. P.).  L. \v C. P. \& E.B. thank to Institute for
Astrophysics Canarias for the hospitality  before and after
observations. Also, we would like to thank  the anonymous referee
for very useful comments.
\end{acknowledgements}

{}

\end{document}